%% file: main.tex
\documentclass[CRPHYS,Unicode,manuscript]{cedram}
\pdfoutput=1
\usepackage{graphicx} 
\usepackage{xcolor} 
\usepackage{amsmath,amsthm,amssymb,amsfonts} 
\usepackage{mathtools}
\usepackage{hyperref} 
\hypersetup{
    colorlinks=true,
    linkcolor=red,
    filecolor=pink,     
    citecolor= magenta,
    urlcolor=cyan
    }
\usepackage{physics} 
\usepackage{subfiles} 

\newcommand{\p}{\partial}

\title{Optomechanical Backreaction of Quantum Field Processes in Dynamical Casimir Effect\texorpdfstring{\\ \small{Invited refereed paper in \protect\emph{\currentjournaltitle}: Special Issue on Analogue Gravity  v9 Aug 7, 2023.}}{}}
\author{\firstname{Yu-Cun} \lastname{Xie}}
\address{Department of Physics, University of Maryland, College Park, MD 20742, USA}
\email[Y. C. Xie]{xyc@terpmail.umd.edu}
\author{\firstname{Salvatore} \lastname{Butera}}
\address{School of Physics and Astronomy, University of Glasgow, Glasgow G12 8QQ, UK}
\email[S. Butera]{salvatore.butera@glasgow.ac.uk}
\author{\firstname{Bei-Lok} \lastname{Hu}}
\address{Maryland Center for Fundamental Physics and Joint Quantum Institute,\\University of Maryland, College Park, MD 20742, USA}
\email[B. L. Hu]{blhu@umd.edu }

\begin{abstract}

\subfile{Abstract}

\end{abstract}

\begin{document}

\maketitle

\section{Introduction}

In keeping with the request from the editors of this special issue dedicated to the memory of Renaud Parentani we have chosen a topic of current research interest which Renaud has worked on and made important contributions, namely, moving mirrors and dynamical Casimir effect (DCE). This ranges from the early substantive papers of Obadia and Parentani \cite{RenaudMM,RenaudMM+,RenaudMM++}, a must-read after that of Davies and Fulling \cite{DavFul,DavFul+}, to the recent papers of Busch, Parentani and Robertson \cite{RenaudDCE,RenaudDCE+}, which use the example of DCE to explore the fundamental issue of quantum entanglement, with the purpose of providing a much more accessible experimental platform than black holes and the early universe for testing the informational theoretical predictions of Hawking-Unruh effects \cite{Haw75,Unr76} and the inflationary cosmology \cite{Guth}. The underlying physical mechanism of DCE, namely, parametric amplification of quantum fluctuations, is a close analogy to cosmological particle creation (CPC) \cite{Par69,Zel70}, a topic studied in the 70s as an important part of the pioneering efforts towards establishing a quantum field theory in curved spacetime \cite{Fulling,Wald,BirDav,ParTom,HuVer}. 

Dynamical Casimir effect has a broader scope than the moving mirror analogue of Hawking radiation, also explored from the 70s on \cite{DavFul,Walker,Carlitz,ChuVer,ChuVer+}, since the trajectory of the mirror is chosen to mock up the position of a black hole (see \cite{GoodWilczek} for varying trajectories) and because of the presence of an event horizon, the particle creation spectrum in this category of problems is specified to be thermal in nature\footnote{Cosmological particle creation can be thermal, the common denominator for thermal particle production in evolutionary cosmology and in spacetimes with event horizons is explained in \cite{KHMR}. One can also treat cases with near-thermal radiation with a stochastic field theory approach, as described in \cite{RHK}.}. Backreaction of Hawking effect using the moving mirror analogy has been studied by some authors earlier on \cite{OkuTsu,Hotta}, including a note by Massar and Parentani  \cite{MasPar}, and more recently, e.g., \cite{Good}. From the rigorous calculations of one of the present authors with Galley and Behunin \cite{GalBehHu},  whereas the late time behavior of the black hole's dynamics as modified by the emitted Hawking radiation is the focus of the black hole backreaction program, it is precisely in the late time regime when divergence problems arise which destroys the moving mirror analogue\footnote{E.g., Haro \cite{Haro} avoided this issue by assuming a well defined out state at late times, thus assuming, rather than predicting, the black hole's eventuality.}.  

Emulating Renaud's style, which is always direct, probing,  serious and thorough, we shall spare the readers with a heavy reference-laden background, leaving it to some excellent recent reviews, e.g.,  \cite{Dodonov,Dodonov+,Ford}, but go directly into the problem and its central issues. Noteworthy is that, behind DCE rests the very important and rapidly developing field of quantum optomechanics (QOM). From the perspective of atomic-optical physics, the displacement of a mechanical membrane have been used to capture the activities in an optical cavity \cite{OMrev,OMrev+}. Reversing the order is the working principle behind LIGO-VIRGO-KAGRA, i.e., using an interferometer aided by optical spring and noise reduction techniques to measure the tiniest displacements of the mirrors in response to impinging gravitational waves. 

\paragraph{Backreaction in 1+1D DCE}
Backreaction effects in physical situations involving moving mirrors have caught increasing attention in the last decades~\cite{KardarRMP1999}, demonstrating vacuum induced dissipation of mechanical motion \cite{Unruh-PRD-2014,Savasta-PRX-2018,Butera-PRA-2019}, as well as fluctuations and quantum decoherence~\cite{Dalvit-PRL-2000,MaiaNeto-PRA-2000,Butera-EPL-2019}.
Since the intent of our present work is to serve more of an illustrative than practical purpose, we shall be using simple prototype models both for 1+1D and 3+1D cases. 

For 1+1D, we consider a ring with a radius which varies with time in response to the quantum effects of the matter field present in the process. The ring can also be understood as two point mirrors at a distance of $L$ apart, being identified with a $S^1$ topology, whereby $L$ becomes the circumference of the ring. (The mirrors need be perfectly conducting to facilitate a reflective periodic boundary condition.) Since the model is in 1+1D, and quantum fields in 1+1D are conformal, the only contribution to DCE is from the conformal anomaly. (In fact, it had been shown \cite{ChrFul77} that the trace anomaly fully determines the Hawking effect in 1+1D black holes.) We shall perform an adiabatic regularization of the energy density of the quantum field, that we use to derive the action for the dynamics of the ring-field system, and eventually the equation of motion of the ring.  In the appendix, we show an alternative derivation of the equation of motion of the ring, which shows explicitly how the trace anomaly appears. Solving this equation gives us the backreaction of quantum matter field, here in the form of the conformal anomaly, on the ring's dynamics.

While DCE has been studied by many authors, using simple models like two parallel conducting plates with time-varying separation \cite{ParallelPlateDCE1,ParallelPlateDCE2}, we know of only one earlier paper \cite{2J} which studied the model of a ring in a breathing mode.  These authors used a path integral (Polyakov) method (see also \cite{GolKar,WuLee,WuLee+,Fosco,Fosco+}) to derive the regularized effective action. Using a different method we obtained the same equation of motion as in \cite{2J}, where these authors calculated the force due to DCE and compared it with the static Casimir force.  They did not address fully the backreaction effects of DCE, meaning, how the radius changes in time in response to the changing backreaction of the quantum field due to the trace anomaly.   

\paragraph{Backreaction in 3+1D DCE and Cosmology}
Regarding quantum effects of matter fields in a 3+1D background, for a conformal field in a conformally-flat spacetime, e.g., the Friedmann-Lem\^aitre-Robertson-Walker (FLRW) universe, the only source in the backreaction equation is also the trace anomaly (see, e.g., \cite{FHH79}). When one considers nonconformal fields in a conformally-flat spacetime (e.g., gravitons in FLRW universe \cite{Har77,HuPar77}) or conformal fields in a non-conformally flat spacetimes (e.g., the Bianchi type I universe \cite{HuPar78,HarHu79,HarHu80,CalHu87}, or inhomogeneities in a FLRW universe \cite{CamVer,CamVer+})  there will be particle creation and the backreaction would come from two sources, one due to the trace anomaly (attributable to vacuum polarization effects), the other due to particle creation originating from the parametric amplification of vacuum fluctuations (spontaneous creation) or from particles already present (stimulated creation) \cite{HKM94}.  It has been shown that particle creation at the Planck time  led to a rapid isotropization and homogenization of the early universe. For backreaction effects of quantum fields on the singularity and particle horizon issues,  see \cite{And83}.

\paragraph{Analogy: Similarities and Differences}
Despite the soundness of this analogy, when it comes to backreaction effects, there is an important basic difference between DCE and cosmological particle creation (CPC).  In DCE, there is an external agent which drives the ring or the mirror. How much energy this agent puts into driving the system, resulting in particle creation of varying degrees,  is determined by the agent. If one keeps pumping energy into the system at a steady rate, particle creation will continue and its backreaction effect on the external agent would be marred by the stipulated drive protocol (albeit the agent should feel its reactive force).  In cosmology, the agent is the universe, but it does not have a free reign. As we know from classical general relativity, the Einstein equation is both an equation of motion, dictating how matter moves or behaves in a curved spacetime; it is also a field equation for the dynamics of spacetime -- matter governs how space is curved and evolves. Same situation in the semiclassical Einstein equation, where matter is described by quantum fields. Spacetime and matter make up a closed system, and the interplay between these two sectors has to satisfy a self-consistency condition. This is explicitly demonstrated in \cite{CalHu87} where the energy density of particles produced over time in the field sector is shown to be exactly equal to the dissipative energy (anisotropy damping) in the geometrodynamics sector.   Therefore, to make a sound comparison of backreaction effects between DCE and CPC while solving for the dynamics of the ring / mirror it is important to place a constraint by specifying a fixed total energy of the ring/mirror - quantum field closed system.    

\paragraph{Quantum Lenz Law} 
Given this condition we expect the central feature of backreaction of particle creation in DCE would be similar qualitatively to that of backreaction in cosmological particle creation, namely, obeying the `Lenz Law', termed by the leading practitioners in 80s \cite{Hu83,Par83,Mot86,CalHu89}, which says that the backreaction effect of particle creation on the drive (in cosmology it is the evolution of the universe, in DCE, the external agent, but note the necessity of imposing the condition of constant total energy) works in such a way as to resist the changes, i.e., a reduction in the production of particles and a slowdown in the motion of the mirror. In the example of particle creation from a conformal quantum scalar field in an anisotropically expanding universe or a universe with small inhomogeneity,   this means the universe would eventually (indeed very quickly near the Planck time) become isotropic and homogeneous whence particle creation ceases. 

\paragraph{Key Findings} Indeed this is what we found in the 3+1D case, for a conformal scalar field in a rectangular metal (or mirror with perfect reflection) box with one face free to move determined only (no external agent) by the backreaction effects of particle creation.  Note that particles are produced both when the ring expands or contracts. Our backreaction results show that particle creation diminishes and the mirror slows down in both cases.  What is intriguing is backreaction effects in the 1+1D case,  where there is no particle creation and conformal anomaly is the only quantum field process present. The static Casimir force in our setup (a ring can be seen as two point mirrors being identified as one) is attractive thus there is already the tendency for the ring to shrink. The rather curious result for the backreaction of trace anomaly is that it accelerate the contraction -- something interesting enough for further thoughts.

\section{Backreaction in 1+1 dimension}

\subfile{2d}

\section{Backreaction in 3+1 dimension}

\subfile{4d}

\section{Related Problems for Further Development}

\subfile{RP}

\appendix
\section{Calculation of trace anomaly in 1+1D}

\subfile{appendix1}

\bibliographystyle{crunsrt}
\bibliography{refs.bib}
\end{document}

%% file: Abstract.tex
In the last two decades researchers in analogue gravity have skillfully designed many laboratory accessible experiments to test some important effects in semiclassical gravity, fluids and BEC analogue of Hawking-Unruh radiation being  a prominent example. The analogue of cosmological particle creation is dynamical Casimir effect, as they share the same underlying physical mechanism, that of parametric amplification of vacuum fluctuations in the quantum field by an expanding universe or by a fast moving boundary. Backreaction of cosmological particle creation at the Planck time has been shown to play a significant role in the isotropization and homogenization of the early universe. The corresponding problem of understanding the backreaction effects of quantum field processes in DCE is the goal of this work.  We present analyses of quantum field processes in two model systems: in 1+1D, a ring (made up of two point mirrors identified at the two ends) with time-dependent radius, and in 3+1D,   a symmetric rectangular conducting box where the length of one side is allowed to change with time.  In both cases the time-dependence of the radius or the length is not prescribed -- no external agent present --  but determined solely by the backreaction of particle creation and related effects. We find that for 1+1D, the only quantum field effect due to the trace anomaly tends to accelerate the contraction of the ring over and above that already present due to the weaker attractive force in the static Casimir effect. For the rectangular box, the rate of expansion or contraction of the box is slowed down compared to that due to the static Casimir effect. Our findings concur with results found in cosmological backreaction problems, which can be encapsulated into what is known as the quantum Lenz law,  that the backreaction works in the direction of opposing further changes, which means the suppression of particle creation and a slow down of the system dynamics. In conclusion we also mention two related classes of problems of theoretical significance for further investigations. 


%% file: 2d.tex
In this section we work with a simple 1+1D geometry, in which the backreaction of a massless conformal quantum field is due exclusively to the trace anomaly. In the next section, we consider a 3+1D dimensional configuration and highlight the other kind of backreaction, due to particle creation.   

For simplicity, let us consider a spacetime manifold with topology $\mathbf{R}\times S^1$, that is a circular one-dimensional ring  $S^1$, with circumference (coordinate length)  $l$. We allow the physical length $L(t)\equiv a(t)l$ of the ring to change in time with a time-dependent scale factor $a(t)$. 
The line element $ds$ of this 1+1D spacetime is given by ($c=\hbar=1$):
\begin{equation}
    ds^2 \equiv g_{\mu\nu} \dd x^\mu \dd x^\nu = \mathrm{d}t^2 - a(t)^2 \mathrm{d}x^2
    \label{metric},
\end{equation} 
where $g_{\mu\nu}$ is the metric tensor.
Let us consider a massive $m$ conformally-coupled scalar field whose action, $S_f[\phi,g_{\mu\nu}]$, has the form:
\begin{equation}
	S_f [\phi,g_{\mu\nu}] = \int \dd t \int_0^l \dd x \,|g|^{1/2}\frac{1}{2}\Big[g^{\mu\nu}\big(\partial_\mu\phi\big)\big(\partial_\nu\phi\big)-m^2\phi^2\Big].
\label{Eq:Sf}
\end{equation}
Here, $g \equiv \text{det}[g_{\mu\nu}] = -a^2$, while $g^{\mu\nu}$ is the inverse metric tensor: $g^{\mu\alpha}g_{\alpha\nu} = \delta_\nu^\mu$.

Our objective is to compare how the scale factor changes with time in the two cases: 1) Due solely to the Casimir effect (no backreaction), 2) due to the combination of Casimir effect and backreaction from the trace anomaly of the conformal field.  Note that, in the first case, the size of the ring decreases due to the {\it attractive} Casimir force. 
We thus want to find out how the backreaction of quantum field in the form of the trace anomaly may affect the rate of change of the radius -- does it slow down the shrinkage or speed it up? 

To this end, we derive in what follows an effective action for the dynamics of the ring, where the energy density of the field enters in the form of extra kinetic and potential terms. As usual,  in quantum field theory, the \emph{bare} energy density suffers from ultraviolet divergences due to the presence of zero-point fluctuations, which needs to be regularized or renormalized (if the divergences can be put in the form of covariant geometric quantities) in order to obtain a finite \emph{physical} energy density. We perform this task in the next section.

\subsection{Adiabatic Vacuum and Regularization}

There are many ways to first identify and then subtract out the ultraviolet divergences. For time-dependent configurations as the one we are considering, adiabatic regularization is probably the most intuitive and easy to use. At the time when the original papers \cite{ParFul74,FulPar74,FulParHu74} introducing this method were written (see also the related n-wave regularization \cite{ZelSta71} and iterative-time \cite{Hu74} methods) there was no knowledge of the existence of the trace anomaly \cite{CapDuf,Duff}. It was derived later for scalar fields in Bianchi type I \cite{Hu78} and Robertson-Walker spacetimes by Hu \cite{Hu79}, who pointed out the trace anomaly appears only when one works with a massive field and after all the calculations are done, at the end, set mass to zero. The remainders in the reduction of the integrals for higher order divergences to a lower order add up to the expression of the trace anomaly.   Anderson \& Parker \cite{AndPar} pointed out the correct measure to use for closed  Robertson-Walker spacetimes. Later, Navarro-Salas and co-workers \cite{ValenciaFermi,ValenciaYukawa,ValenciaDirac} systematically derived the trace anomalies for Fermi, Dirac and Yukawa fields, and Chu \& Koyama for gauge fields \cite{ChuKoy}, to name a few representative work. Here, to get the trace anomaly we follow Hu's dictum in first calculating the energy-momentum tensor of massive fields and only at the end set the mass equal to zero.

By minimizing the action in Eq.~\eqref{Eq:Sf} with respect to the field's variations: $\delta S/\delta \phi = 0$, and by using the metric in Eq.~\eqref{metric}, we obtain the equation of motion of the field in the form:
\begin{equation}
	\partial_t^2\phi\left(\frac{\dot{a}}{a}\right)\partial_t\phi-\frac{1}{a^2}\partial_x^2\phi + m^2\phi = 0.
\label{Eq:Field}
\end{equation}
Since the system at hand is invariant upon translations, the wave vector $k$ is conserved and the general solution to Eq.~\eqref{Eq:Field} can be taken in the form:
\begin{equation}
	\phi = \sum_k \frac{1}{\sqrt{la}} e^{ikx}\;f_k(t).
\label{Eq:fk}
\end{equation}
Note that, under the specified spacetime topology, the wave vectors $k_n$ that satisfy periodic boundary conditions form a discrete set $k_n=2\pi n/l$, with $n=0,\pm 1,\pm 2,\cdots$ (we drop the subscript \emph{n} for brevity, in what follows). By inserting Eq.~\eqref{Eq:fk} into Eq.~\eqref{Eq:Field}, we obtain that the time-dependent component of the eigenmode solution satisfies the following harmonic oscillator equation:
\begin{equation}
	\ddot{f}_k + w_k^2(t) f_k = 0,
\label{Eq:hk}
\end{equation}
where the frequency $w_k(t)$ depends on time according to:
\begin{equation}
	w_k^2(t) = \omega_k^2 + \sigma(t),
\label{Eq:Omega}
\end{equation}
with
\begin{equation}
    \omega_k^2(t) = \frac{k^2}{a^2} +  m^2,  \qquad   \sigma(t) = -\frac{1}{2} \left(\frac{\ddot{a}}{a} - \frac{\dot{a}^2}{2a^2}\right).
\end{equation}
First,  the vacuum of a quantum field in a static spacetime need be replaced by one suitable for fields in a dynamical setting.  The adiabatic vacuum (of a certain order) can be defined by performaing an adiabatic expansion (i.e., WKB approximation) of the eigenmodes (to that order). To this end, we insert in Eq.~\eqref{Eq:hk} the ansatz
\begin{equation}
	f_k(t) = \frac{1}{\sqrt{W_k(t)}}\exp\left[-i\int^t \dd t'\,W_k(t')\right],
\label{Eq:WKB}
\end{equation}
obtaining the approximated solution for $W_k(t)$, in the form:
\begin{equation}
	W_k^2 = \omega_k^2 - \frac{1}{2} \left(\frac{\ddot{W}_k}{W_k}-\frac{3}{2}\frac{\dot{W}_k^2}{W_k^2}\right).
\label{Eq:Weq}
\end{equation}
An adiabatic vacuum (of a certain order) is defined under the assumption of sufficiently slow time variation of the scale factor $a(t)$ compared to the characteristic frequency of the mode, to ensure there is minimal particle creation in that mode (to that order).    
In the present setting the adiabatic vacuum (of second order) is defined under the condition that the ring evolves on a time scale $T$ much longer than the period of the modes considered. 
By pursuing a perturbative expansion with respect to the small dimensionless parameter $\epsilon \equiv (\omega_k T)^{-1}$, we obtain:
\begin{equation}
    W_k(t) = \omega_k - \frac{1}{4\omega_k}\left(\frac{\ddot{\omega}_k}{\omega_k}-\frac{3}{2}\frac{\dot{\omega}_k^2}{\omega_k^2}-2\sigma\right) + \mathcal{O}(\epsilon^4)
\end{equation}
Together with Eq.~\eqref{Eq:WKB}, the expression for $W_k(t)$ here obtained defines the adiabatic expansion of the field's modes up to second order. By using the basis composed by these modes, the field operator $\hat{\phi}(x,t)$ can then be decomposed as:
\begin{equation}
	\hat{\phi}(x,t) = \frac{1}{\sqrt{la}}\sum_{k= - \infty}^{+\infty}\left[\hat{A}_k f_k(t)e^{ikx} + \hat{A}_k^\dag f_k^*(t) e^{-ikx}\right].
\label{Eq:Qfield}
\end{equation}
The ladder operators $\hat{A}_k$, $\hat{A}_k^\dag$ annihilate and create a particle with wave vector $k$, and define the adiabatic vacuum state of the field as: $\hat{A}_k \ket{0_A} = $ $\forall\,k$.

\subsection{Regularized Energy Density}

The energy density of a massive conformal scalar field in 1+1D spacetime with metric Eq.~\eqref{metric}, is given by:
\begin{equation}
     \rho(x)  =T_{00}= \frac{1}{2}\bigg[ \big(\partial_{t}\phi\big)^2 +  \frac{1}{a^2}\big(\partial_{x}\phi\big)^2+m^2\phi^2\bigg].
\label{Eq:EnDens1D}
\end{equation}
Here, the continuum limit $l\to\infty$ is taken by turning the sum over wave-vectors into the corresponding integrals according to the standard prescription: $\sum_k \to (l/2\pi)\int \dd k$.
By using the adiabatic expansion for the modes previously introduced, the energy density can be expanded up to second order in the adiabatic parameter $\epsilon$, as $\rho(x) = \int_{-\infty}^\infty dk\,\left[\rho_k^{(0)}+\rho_k^{(2)}+\mathcal{O}(\epsilon^4)\right]$, where:
\begin{align}
	\rho_k^{(0)} &= \frac{\omega_k}{4\pi a},\\
	\rho_k^{(2)} &= \frac{1}{8\pi a}\bigg[\left(\frac{\dot{a}}{2a}\right)^2\frac{1}{\omega_k} + \frac{\dot{a}}{2a}\frac{\dot{\omega}_k}{\omega_k^2}+\frac{1}{4}\frac{\dot{\omega}_k^2}{\omega_k^3}\bigg].
	\end{align}
Here, we indicated by $\rho_k^{(n)}$ the contribution to the energy density of order $n$ in the adiabatic expansion. The first term diverges at high momenta, as $\rho_k^{(0)}\sim k^2$, while the second term converges to the finite value
\begin{equation}
	\rho^{(2)} = \int_{-\infty}^\infty dk\,\rho_k^{(2)} = \frac{1}{24\pi} \frac{\dot{a}^2}{a^2}.
\end{equation}
The energy density is regularized by subtracting $\rho^{(0)}$ and $\rho^{(2)}$ from the exact expression for the energy density $\rho_{\rm ex}$, that can be readily calculated for a conformally invariant scalar field and for arbitrary functions of the scale factor: $\rho_{\rm ex} = \sum_{k=0}^{+\infty}{\omega_k}/{(a^2l)}$. That is: $\rho_{\rm reg} = \rho_{\rm ex} - \rho^{(0)} - \rho^{(2)}$.  The first two terms give rise to the static Casimir energy density $\rho_{\rm Cas}$, that is responsible for the appearance of the Casimir force. By introducing a cutoff function to regularize the diverging series and integral, which is removed at the end of the calculations, and by taking the massless limit $m\to 0$, this takes the standard form:
\begin{equation}
\begin{split}
    \rho_{\rm Cas} &\equiv \rho_{\rm ex} - \rho^{(0)}\\
    &= \lim_{\lambda\to 0}\frac{1}{al} \bigg[\sum_{k=0}^{+\infty}\frac{\omega_k}{a} \exp(-\lambda \omega_k/a) - \frac{1}{2\pi}\int_0^{+\infty} dk \, \frac{\omega_k}{a}\exp(-\lambda \omega_k/a)\bigg]\\
    &= -\frac{\pi}{6 a^2 l^2}.
\end{split}
\end{equation}
With this result, we obtain the regularized expression for the energy density of the massless field:
\begin{equation}
	\rho_{\rm reg} = -\frac{\pi}{6 a^2 l^2} - \frac{1}{24\pi}\frac{\dot{a}^2}{a^2}.
	\label{Eq:T00_ren}
\end{equation}
The total energy of the field is finally obtained by integrating the energy density over all space:
\begin{align}
    H &=  \int_{0}^{l}\sqrt{-g_{xx}\dd x^2}\;\rho_{\rm reg} = -\frac{1}{24\pi}\frac{\dot{L}^2}{L}-\frac{\pi}{6L},
    \label{Eq:H}
\end{align}
where we restored the physical length $L(t) = a(t) l$ of the ring.

\subsection{Effective Action and the Equations of Motion}

The effective action for the ring is constructed by incorporating the energy $H$ of the field into the action of the mirror, which we take as a non-relativistic particle of mass $M$. Specifically, we add the term in Eq.~\eqref{Eq:H} involving time derivatives as an kinetic term, whose physical meaning is a ring's size dependent renormalization of the bare mass $M$, and the Casimir energy as a potential term:
\begin{equation}
    S_{\rm eff}=\int \dd t \left[\frac{1}{2}M\dot{L}^2-\frac{1}{24\pi}\frac{\dot{L}^2}{L}+\frac{\pi}{6L}\right].\label{2d action}
\end{equation}
As described in the beginning of this section, the backreaction effect of quantum field under study is highlighted by comparing the present action containing backreaction effects with the action which yields the motion of the ring only under static Casimir force, namely, by dropping the second term in Eq. \eqref{2d action}.   
\begin{equation}
    S_{\mathrm{Cas}}=\int \dd t \left[\frac{1}{2}M\dot{L}^2+\frac{\pi}{6L}\right].
\end{equation}

\begin{figure}[tbp]
    \centering
    \includegraphics[width=0.45\columnwidth]{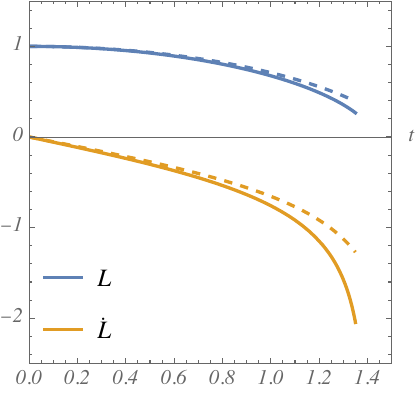}
    \caption{Plot of ring's size and velocity vs. time. The solid represents the motion of the mirror under static Casimir force plus dynamics-induced force, and the dashed line represents the mirror's motion only under static Casimir force.}
    \label{fig:2d dy}
\end{figure}

Assuming the ring is a classical object (no quantum fluctuation), we obtain the equation of motion as the Euler-Lagrangian equation corresponding to the actions:
\begin{align}
	\left(M - \frac{1}{12\pi L} \right) \ddot{L} &= -\frac{\pi}{6 L^2} - \frac{1}{24\pi} \frac{\dot{L}^2}{L^2} &\text{(with bkr)},\label{EqMot_Bkr}\\
	M \ddot{L} &= -\frac{\pi}{6 L^2}  &\text{(no bkr)}.\label{EqMot_NoBkr}
\end{align}
Notice that, in the first equation, part of the effects of backreaction can be understood as inducing a change in the mass of the ring to a lesser effective mass which depends on the length of the ring itself: the smaller the ring, the more significant the effective mass reduction. The second effect of the backreaction is conveyed by the last term on the RHS, which results proportional to the square of the ring velocity. Both the equations of motion \eqref{EqMot_Bkr} and \eqref{EqMot_NoBkr} are numerically solved for $M=1$. The results are shown in Fig. \ref{fig:2d dy}.
We can see that the backreaction effect of quantum field due to the trace anomaly accelerates the collapse of the ring, regardless of the sign of the velocity of the ring $\dot{L}$. This means that the backreaction cannot be understood in terms of a frictional force acting on the ring. Physically, this is due to the absence energy injected into the field due to particle production, which is absent in the 1+1D configuration here considered. As shown in the next section, this physical picture dramatically changes upon considering a more general 3+1 geometry.


%% file: 4d.tex
In this section we consider a massless conformal quantum field in a symmetric (two transverse directions fixed with equal length) rectangular box whose faces are made up of perfectly conducting plates. We allow the length of the third side (longitudinal direction) to change with time. A rapid change will engender particle creation from vacuum fluctuations, which is the more prominent quantum field process in dynamical Casimir effect we wish to focus on here.  We are interested in the backreaction effect of particle creation on the expansion rate of this box. Specifically, as explained in the Introduction,  we want to see if the quantum Lenz law, which we systematized from the backreaction results of cosmological particle creation,  applies to the backreaction effect in the present setup, namely,  the backreaction will act in such a way as to resist further changes, meaning, whether particle creation will slow down the expansion or contraction of the box, and at what rate. 

\subsection{Rectangular mirror box and its cosmological analog}
The rectangular box configuration mimics that of a symmetric Bianchi type-I universe with  metric tensor:
\begin{equation}
    ds^2 = \dd t^2 - \qty[a_1^2(t)\dd x^2 + a_2^2(t)\dd y^2 + a_3^2(t)\dd z^2],
\end{equation}
where $a_i$ are the scale factors in the three directions. We consider the case with $a_2=a_3=1$, $a_1=a(t)$, i.e., with one side moving, in the $\pm x$ direction. To turn this into a box we impose periodic boundary conditions at both ends of each of the three sides, i.e., between $x,y,z = 0$ and $x,y,z = l$, where $l$ is the length.  Our present set up has $l$ fixed in the $y, z$ directions, but allow the scale factor in the $x$ direction to change in time. Assuming conducting boundary surfaces gives us a closed ``mirror box universe'' with topology $\mathbf{R}\times T^3$ where $\mathbf{R}$ refers to the time direction and  $T$ denotes a torus.

Consider a massless conformal scalar field $\phi$ in this box universe obeying the Klein-Gordon equation: 
\begin{align}
    &~~~~~\Box\phi +\frac{1}{6}R\phi = 0,\\
    &\Rightarrow \p_t^2\phi + \frac{\dot{a}}{a}\dot{\phi} - \sum_i\frac{\p_i^2\phi}{a_i^2}+\frac{1}{6}R\phi=0,
\end{align}
where $R$ is the scalar curvature of spacetime.
The conformal energy-momentum tensor of this  scalar field is given by
\begin{align}
    T_{\mu\nu} &= (\p_\mu\phi)(\p_\nu\phi)-\frac{1}{2}g_{\mu\nu}g^{\lambda\sigma}(\p_\lambda\phi)(\p_\sigma\phi) \nonumber\\
    &-\frac{1}{6}\left[\nabla_\mu\p_\nu(\phi^2)-g_{\mu\nu}g^{\lambda\sigma}\nabla_\lambda\p_\sigma(\phi^2)+\phi^2G_{\mu\nu}\right].
\end{align}
We are interested in the  energy density 
\begin{equation}
    T_{00}=\frac{1}{2}(\dot{\phi})^2 + \frac{1}{3}\frac{\dot{a}}{a}\phi\dot{\phi}+\frac{1}{6}\sum_i \left(\frac{\p_i\phi}{a_i}\right)^2-\frac{1}{3}\sum_j \frac{\phi\p_j^2\phi}{a_j^2}-\frac{1}{6}G_{00}\phi^2,
\end{equation}
where an overdot indicates differentiation with respect to $t$.

Let us introduce a new field variable $\chi$ and a conformal time variable $\eta$:
\begin{equation}
    \chi = a^{1/3}\phi, \; \dd\eta =  a^{-1/3}{\dd t}.
\end{equation}
In these variables one gets a conformally-flat metric when the spacetime is isotropic.
By substituting 
\begin{equation}
    \dot{\phi} = a^{-2/3}\left[ \chi' - \frac{1}{3}\frac{a'}{a}\chi\right]
\end{equation} 
(where a prime indicates differentiation respect to $\eta$) into the above equation for the energy density, we have
\begin{equation}
    T_{00}= a^{-2/3}\qty[\frac{1}{2}a^{-2/3}(\chi')^2 +\frac{1}{6}\sum\qty(\frac{\p_i\chi}{a_i})^2-\frac{1}{3}\sum\frac{\chi\p_j^2\chi}{a_j^2}-\frac{1}{2}a^{-2/3}Q\chi^2],
\end{equation}
where  $Q$ is defined as 
\begin{equation}
    Q = \frac{1}{18}\sum_{i<j}\qty(\frac{a_i'}{a_i}-\frac{a_j'}{a_j})^2= \frac{1}{9}\qty(\frac{a'}{a})^2
\end{equation}
representing the degree of anisotropy. Now the scalar field equation becomes 
\begin{equation}
    \chi'' - a^{2/3}\sum \frac{2\p_i^2\chi}{a_i^2}+Q\chi = 0
\end{equation}
Although this spacetime is anisotropic, it remains homogeneous. Under the periodic condition imposed at the boundaries we can perform the following mode decomposition:
\begin{equation}
    \chi = l^{-3/2}\sum_\mathbf{k} \qty[A_{\vec{k}}\chi_{\vec{k}}(\eta)e^{i\vec{k}\cdot\vec{x}}+A_{\vec{k}}^{\dagger}\chi_{\vec{k}}^*(\eta)e^{-i\vec{k}\cdot\vec{x}}]
\end{equation}
with the compact notation $$\sum_{\vb{k}}= \prod_{i=1}^{3}\sum_{n_i}$$
where $k_i=2\pi n_i/l,\; n_i = 0, \pm 1, \pm 2 \pm \cdots$.
The Fourier modes $\chi_{\Vec{k}}$ satisfy the parametric oscillator equation:
\begin{align}
    \chi_{\Vec{k}}''+(\Omega_{\Vec{k}}^2+Q)\chi_{\vec{k}}=0
    \label{xeq}
\end{align}
with time-dependent frequency 
\begin{equation}
    \Omega_{\Vec{k}}^2=\qty(a^{1/3}\omega_{\vec{k}})^2=a^{2/3}\qty(\sum\frac{k_i^2}{a_i^2}+m^2)=a^{2/3}\qty(\frac{k_x^2}{a^2}+k_y^2+k_z^2+m^2).
\end{equation}

Let $\ket{0_A}$ be the vacuum state defined by $A_{\vec{k}}$. The vacuum expectation value of the $00$ component of the stress-energy tensor 
\begin{equation}
    \expval{T_{00}}{0_A}= {\frac{1}{2l^3}}a^{-4/3}\sum_{\vb{k}}\bqty{\vqty{\chi_{\vec{k}}'}^2+(\Omega_{\vec{k}}^2-Q)\vqty{\chi_{\vec{k}}}^2}
    \label{div}
\end{equation}
gives the energy density we want. However, the above expression of energy density diverges due to high frequency (UV) contributions. One can remove the divergences by adiabatic regularization procedures described in \cite{FulParHu74} which we adopt in what follows (subscript $k$ on $\Omega$ are dropped). \footnote{Due to the imposition of boundary conditions the global topology of our setup is not the same as the Bianchi type-I Universe. However, UV divergence is a local property not affected by the global topology of spacetime.}
\begin{align}
    \expval{T_{00}}_{\text{div}}&=\frac{1}{32\pi^3a^{4/3}}\int d^3k\Omega^{-1}\biggl\{\biggl.2\Omega^2+\qty[\frac{1}{4}\qty(\frac{\Omega'}{\Omega})^2-Q]\nonumber\\
    &-\frac{1}{8}\qty(\frac{\Omega'}{\Omega})^2\epsilon_{2(2)}+\frac{1}{4}\frac{\Omega'}{\Omega}\epsilon'_{2(3)}+\frac{1}{4}\Omega^2\epsilon_{2(2)}^2+\frac{1}{2}Q\epsilon_{2(2)}\biggl\}
    \label{divt}
\end{align}
Note that the first term represents divergence already present in flat space and need be subtracted out in all circumstances. 

The regularized energy density from particle creation is obtained by subtracting Eq. \eqref{divt} from Eq. \eqref{div} with 
the following results 
\begin{align}
    \expval{T_{00}}&={\frac{1}{2l^3}}a^{-4/3}{\sum_{\vb{k}}}\bqty{\vqty{\chi_{\vec{k}}'}^2+(\Omega^2-Q)\vqty{\chi_{\vec{k}}}^2}\nonumber\\
    &-\frac{1}{32\pi^3a^{4/3}}\int d^3k\Omega^{-1}\biggl\{\biggl.2\Omega^2+\qty[\frac{1}{4}\qty(\frac{\Omega'}{\Omega})^2-Q]\nonumber\\
    &-\frac{1}{8}\qty(\frac{\Omega'}{\Omega})^2\epsilon_{2(2)}+\frac{1}{4}\frac{\Omega'}{\Omega}\epsilon'_{2(3)}+\frac{1}{4}\Omega^2\epsilon_{2(2)}^2+\frac{1}{2}Q\epsilon_{2(2)}\biggl\}.
    \label{fullEnergy}
\end{align}

\subsection{Regularized Energy Density}

The expression of energy density obtained in the previous section can be divided into three parts:
\begin{align}
    \expval{T_{00}}&=\rho_\text{creation}+\rho_\text{matter}+\rho_\text{Casimir},
\end{align}
where  $\rho_\text{create}$ refers to the contribution from freshly created particles at the instant and vacuum polarization effects induced by spacetime dynamics, $\rho_\text{matter}$ from accumulated created particles in the past forming a relativistic fluid of conformal matter (like photons),  and $\rho_\text{Casimir}$ from the static Casimir energy density because of the boundary conditions we imposed. If we assume the initial state is the vacuum state, the matter contribution to the energy density is initially zero. As time evolves and particle creation occurs, the effervescent creation component will continue to flow into the matter contribution. This division is inspired by the cosmological backreaction considerations in \cite{HuPar77,HuPar78}. There, the quantum effects of particle creation was found to dominate over all other contributions to the total energy density. We expect a similar situation in the present problem.  

To find the backreaction of particle creation, we shall focus on the creation component contribution to energy density $\rho_\text{creation}$.  We shall give an estimate of the matter component, which indeed turns out to be orders of magnitude weaker.  We shall also ignore the Casimir contribution. This can be done by taking the limit $l\rightarrow\infty$ in Eq. \eqref{fullEnergy}, resulting in 
 
\begin{align}
    \rho_\text{creation}+\rho_\text{matter}&=\frac{1}{16\pi^3a^{4/3}}\int d^3k\bqty{\vqty{\chi_{\vec{k}}'}^2+(\Omega^2-Q)\vqty{\chi_{\vec{k}}}^2}\nonumber\\
    &-\frac{1}{32\pi^3a^{4/3}}\int d^3k\Omega^{-1}\biggl\{\biggl.2\Omega^2+\qty[\frac{1}{4}\qty(\frac{\Omega'}{\Omega})^2-Q]\nonumber\\
    &-\frac{1}{8}\qty(\frac{\Omega'}{\Omega})^2\epsilon_{2(2)}+\frac{1}{4}\frac{\Omega'}{\Omega}\epsilon'_{2(3)}+\frac{1}{4}\Omega^2\epsilon_{2(2)}^2+\frac{1}{2}Q\epsilon_{2(2)}\biggl\}.
    \label{noCasimirEnergy}
\end{align}

To select out the creation contribution to the energy density we need to identify the region in $k$-space which yields the most particle production. They are the modes which are subjected to strong nonadiabatic parametric amplification. For each $\bf k$ mode one can use the nonadiabaticity parameter introduced in  \cite{Hu74} (Eq.73-74):  $\bar \Omega \equiv \Omega'_{\mathbf{k}}/\Omega^2_{\mathbf{k}}$ in conformal time, or $\bar \omega \equiv \dot\omega_{\mathbf{k}}/ \omega^2_{\mathbf{k}}$ in cosmic time (Minkowski time here) to measure how rapidly it changes in response to the changing background through the scale factor $a(\eta)$ or $a(t)$. Those fulfilling the condition $\bar \Omega \gg 1$ in conformal time, or $\bar \omega \gg 1$ in cosmic time, are the modes which contribute most to particle creation.  Integrating over these highly non-adiabatic modes in Eq. \eqref{noCasimirEnergy} will capture the dominant contribution to the energy density due to particle creation. 

Examining this condition further, we see that 
\begin{equation}
    \frac{\dot \omega_{\mathbf{k}}}{\omega^2_{\mathbf{k}}} \gg 1   \Rightarrow  \frac{\p\omega_{\mathbf{k}}}{\omega_{\mathbf{k}}} \gg \omega_{\mathbf{k}} \p t     \Rightarrow 
    \text{dimensionless quantity} \gg \omega_{\mathbf{k}} t.\label{eq:nonac}
\end{equation}
The Eq. \eqref{eq:nonac} says that the nonadiabaticity condition is satisfied by the approximation that $\omega_{\mathbf{k}}t \ll 1$. This corresponds to an ellipsoid region in $k$-space which we denote 
by $\mathcal{R}(t)$.

\subsection{Particles created from nonadiabatically amplified modes}

Now, within the region $\mathcal{R}(t)$, the equation of $\chi_k$ \eqref{xeq} can be solved under the low-$k$ or early-$t$ approximation \cite{ZelSta71,HuPar77,HuPar78}.

First, one reduces Eq. \eqref{xeq} into two first-order differential equations by the introduction of two arbitrary functions $\alpha_k$ and $\beta_k$:
\begin{equation}
    \chi_{\vec{k}}=(2\Omega_k)^{-1/2}\qty[\alpha_k e^{-i\int^{\eta}\Omega_k\dd\eta'}+\beta_k e^{i\int^{\eta}\Omega_k\dd\eta'}],
    \label{chi1}
\end{equation}
\begin{equation}
    \chi'_{\vec{k}}=-i(\Omega_k/2)^{1/2}\qty[\alpha_k e^{-i\int^{\eta}\Omega_k\dd\eta'}-\beta_k e^{i\int^{\eta}\Omega_k\dd\eta'}].
    \label{chi1d}
\end{equation}
The Wronskian    condition $\chi'^*_k\chi_k-\chi_k^*\chi'_k=i$ becomes 
\begin{equation}
    \qty|\alpha_k(\eta)|^2 - \qty|\beta_k(\eta)|^2=1.
\end{equation}
Combining Eq. \eqref{xeq}, \eqref{chi1} and \eqref{chi1d}, we obtain two first-order ODE derived from the second order Eq. \eqref{xeq}.
\begin{equation}
    \alpha'=\frac{1}{2}\qty(\frac{\Omega'}{\Omega}-i\frac{Q}{\Omega})\beta e^{2i\int^{\eta}_{\eta_0}\Omega\dd\eta'}-i\frac{Q}{2\Omega}\alpha,
\end{equation}
and
\begin{equation}
    \beta'=\frac{1}{2}\qty(\frac{\Omega'}{\Omega}+i\frac{Q}{\Omega})\alpha e^{-2i\int^{\eta}_{\eta_0}\Omega\dd\eta'}+i\frac{Q}{2\Omega}\beta.
\end{equation}
Under the low-frequency and early time approximation $\omega_{\mathbf{k}}t < 1$, an integral like $\int_{\eta_0}^{\eta}\Omega\dd\eta'$ can be approximated to $0$. Thereby we can find the general solutions of $\alpha$ and $\beta$:
\begin{align}
    \alpha&=c_1\qty(\Omega^{1/2}-i\Omega^{-1/2}\int^{\eta}_{\eta_0}Q\dd\eta')+c_2\Omega^{-1/2},\\
    \beta&=c_1\qty(\Omega^{1/2}+i\Omega^{-1/2}\int^{\eta}_{\eta_0}Q\dd\eta')-c_2\Omega^{-1/2},
\end{align}
where $c_1(k)$ and $c_2(k)$ are complex numbers satisfying 
\begin{equation}
    c_1c_2^* + c_2c_1^* = \frac{1}{2}
\end{equation}
under the normalization condition. With this, Eq. \eqref{noCasimirEnergy} can be solved for $\chi_k$:
\begin{align}
    \vqty{\chi_{\vec{k}}'}^2+(\Omega^2&-Q)\vqty{\chi_{\vec{k}}}^2=2|c_1|^2\qty[(\Omega^2-Q)+\qty(\int_{\eta_0}^{\eta}Q\dd\eta')^2]\nonumber\\
    &+2|c_2|^2+2i(c_1^*c_2-c_1c_2^*)\qty(\int_{\eta_0}^{\eta}Q\dd\eta').
\end{align}


Regarding the initial conditions,  by choosing the state which corresponds to the absence of particles at time $t_0$, we can impose the initial condition
\begin{equation}
    \alpha_k(t_0) = 1,\;\beta_k(t_0) = 0,
\end{equation}
the values of $c_1$ and $c_2$ are uniquely determined: 
\begin{equation}
    c_1 =\frac{1}{2}\Omega_0^{-1/2}, \; c_2 = \frac{1}{2}\Omega_0^{1/2}
\end{equation}
where $\Omega_0(k)$ is the value of $\Omega_k$ at $t_0$. 

\subsection{Energy density from particle creation} 
We  obtain an expression of $\rho_\text{creation}$ in term of $a$:
\begin{align}
    \rho_\text{creation}&=\frac{1}{8\pi^3a^{4/3}}\int\displaylimits_{\mathcal{R}(t)} d^3k\Biggl\{ \Biggr. \qty[\frac{(\Omega^2-Q)}{\Omega_0}+\frac{\qty(\int_{\eta_0}^{\eta}Q\dd\eta')^2}{\Omega_0}]
    +\Omega_0\Biggl.\Biggr\}-\rho_{\text{div}}
\end{align}

Notice that the divergent part of the energy density Eq. \eqref{divt} obtained from adiabatic regularization should apply to the whole $k$-space. However, to obtain the $\rho_\text{creation}$, we need only consider the region $\mathcal{R}(t)$ in $k$-space. This low frequency / short time regime is opposite to the regime where UV divergences arise and where regularization schemes are designed for their removal. For this reason  there is no need to subtract out more divergent terms than those which are required for field theories in Minkowski space, namely, the $2\Omega_k$ term.  Otherwise, one would get negative contributions because in $\mathcal{R}(t)$ there wasn't any high frequency mode contributions to the energy density to begin with.  

Also, examining the term containing $Q$ integrated over time, the range of integration covers the moment {\it after} particle creation. In the way we separate the three kinds of contributions, this belongs to the energy density attributed to matter and thus should be added to the matter part, which we shall show is quite insignificant compared to that from particle creation.

\subsubsection{\texorpdfstring{$\rho_\text{creation}$}{creation} is the dominant contribution}

With these considerations we get an expression for  $\rho_\text{creation}$:

\begin{align}
    \rho_\text{creation}&=\frac{1}{8\pi^3a^{4/3}}\int\displaylimits_{\mathcal{R}(t)} d^3k\Biggl\{ \Biggr. \frac{(\Omega^2-Q)}{\Omega_0}
    +\Omega_0-2\Omega\Biggl.\Biggr\}
\end{align}
This integration can be explicitly carried out. We first define $k_{yz} = \sqrt{k_y^2+k_z^2}$, and set $a(t_0)=1$. The domain of integration $\mathcal{R}(t)$ is an ellipse in $k$ space:
\begin{align}
    (k_{yz}t)^2+(k_xt/a)^2 \leq 1\\
    \Rightarrow k_{yz}^2+\qty(\frac{k_x}{a})^2 \leq \frac{1}{t^2}.
\end{align}
Recognizing the axial symmetry in the problem, we employ cylindrical coordinates and find:
\begin{align}
\rho_\text{creation}&=(576 \pi ^2 a^{10/3} t^4)^{-1}\Biggl[ \Biggr.9 a^4-36 a^{10/3}+18 a^{8/3} \mathcal{P} +9 a^2 \mathcal{P}-4 a'^2 \mathcal{P} t^3\Biggl.\Biggr],
\end{align}
where 
\[\mathcal{P}=
\begin{cases}
 \frac{\sin ^{-1}\left(\sqrt{a^{-2}-1} a\right)}{\sqrt{a^{-2}-1}} & a<1 \\
 \frac{\log \left(\left(\sqrt{1-a^{-2}}+1\right) a\right)}{\sqrt{1-a^{-2}}} & a>1 \\
 1 & a=1
\end{cases}.\]


We can construct an action for the dynamics of this moving mirror system by placing the $\rho_\text{creation}$ and $\rho_\text{matter}$ as the kinetic terms, and $\rho_\text{Casimir}$ as a potential term:
\begin{align}
    S&= \int \dd t \qty{\frac{1}{2}m\dot{L}^2 + \int\displaylimits_{\text{box}} d^3x \qty[\rho_\text{creation}+\rho_\text{matter}-\rho_{\text{Casimir}}]}\\
    &= \int \dd t \qty{\frac{1}{2}m\dot{L}^2 + Ll^2 \qty[\rho_\text{creation}+\rho_\text{matter}-\rho_{\text{Casimir}}]}\\
    &= \int \dd t \qty{\frac{1}{2}m\dot{L}^2 + E_\text{creation}+E_\text{matter}-E_{\text{Casimir}}}
\end{align}
where $L \equiv al$ is the time-dependent longitudinal length of the box.

\subsubsection{\texorpdfstring{$\rho_\text{Casimir}$ and $\rho_\text{matter}$}{casimir and matter} contributions are much weaker}

We know the $\rho_\text{Casimir}$ generates an attractive force and only depends on the side length of the box. The explicit form of $\rho_\text{Casimir}$ is given in e.g.,\cite{MamaevTrunov}. Although $\rho_\text{Casimir}$ has an effect over time, it is not significant enough to impact the dynamics of the box in a short amount of time. This is because particle creation at an early stage dominates over the effect of $\rho_\text{Casimir}$.
So we shall ignore it in our present consideration.

As for $\rho_\text{matter}$, one may expect its contribution acts like radiation pressure. So let us make an estimate of its effect.

As explained earlier, particle creation is the strongest in those (fast changing) modes which are subjected to nonadiabatic parametric amplification. The remaining modes in the $k$-space are only weakly excited and can be described by using an adiabatic approximation. The $\rho_\text{matter}$ term actually contains both particles in these adiabatic modes which are exponentially small as well as the already created particles being red (blue) shifted as the box expands (contracts). These two components comprise a classical relativistic fluid. Their cosmological backreaction effects have been studied by Lukash and Starobinsky \cite{LukSta}. The overall effect is much weaker than particle creation from the vacuum,  understandably so, as the latter is of nonadiabatic origin. 

The energy density attributed to classical matter under these assumptions is given by  
\begin{equation}
    \rho_\text{matter}= \frac{1}{(2\pi)^3a}\int\displaylimits_{\text{$\mathcal{C}$}}d^3k\; \omega_k \;\abs{\beta_k}^2
\end{equation}
where  $\mathcal{C}$ denotes the complement of $\mathcal{R}$ in $k$-space. This equation is not easy to solve since the domain of $\mathcal{C}$ changes over time. 
However, we can estimate its contribution by the following argument: 
Assuming the box is a closed system, we require the total energy of this system to be conserved, i.e., the total derivative of energy with respect to time equal to $0$. Then we get an equation containing $\dot{E}_\text{matter}$. We can use the initial condition at $t_0$ to solve the value of $\dot{E}_\text{matter}$ near the initial time. Using this value to estimate the cumulative value of $E_\text{matter}$ through the range of time we are interested in will give the maximum possible value of $E_\text{matter}$ because it is linked to the particle creation effect which is the strongest near the initial time. This value is still $10$ orders of magnitude smaller than the rest of the energy. Therefore, neglecting the $\rho_\text{matter}$ term in the action is reasonable.


\subsubsection{Results: Quantum Lenz Law Observed}
\begin{figure}[tbp]
    \centering
    \includegraphics[width=0.45\columnwidth]{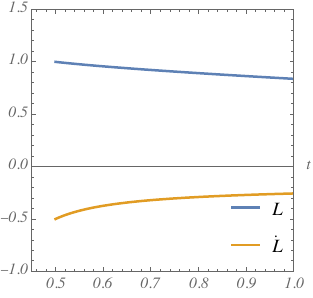}
    \includegraphics[width=0.45\columnwidth]{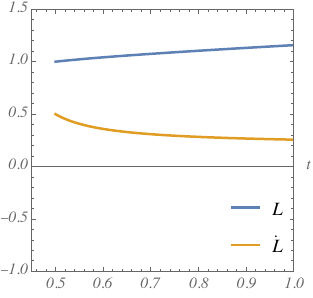}
    \caption{Plot of mirror's position and velocity vs. time. Left figure for $\dot{L}(t_0)=-0.5$, right figure for $\dot{L}(t_0)=0.5$.}
    \label{fig:4d dy}
\end{figure}

\begin{equation}
    S= \int \dd t \qty{\frac{1}{2}m\dot{L}^2 + E_\text{creation}}
    \label{box action}
\end{equation}
Assuming that the moving mirror is a classical object, taking the functional variation of the action in Eq. \eqref{box action} gives the Euler-Lagrange equation for the dynamics of the mirror system. This equation of motion is numerically solved for cases when the mirror initially is left moving (expanding box) and ii) when the mirror is initially right moving (shrinking box). We chose parameters $l=50$ and $m=10$.  The results are shown in Fig. \ref{fig:4d dy} 

In summary, combining the results of the 1+1D case where the backreaction comes solely from conformal anomaly and the 3+1D case where the backreaction comes primarily from particle creation, we can make these observations: 

1. Backreaction from particle creation (from our 3+1D results) obeys quantum Lenz law. This complies with the results from many studies of backreaction effects in cosmological particle creation. 

2. Backreaction effects due to the trace anomaly seem to be very different from that due to particle creation.  This is not surprising because a marked difference also shows up in the cosmological setting if we compare the effect of the trace anomaly in \cite{FHH79} with that of particle creation in \cite{HarHu79}. One can attribute this to the difference between effects originating correspondingly from vacuum polarizations versus vacuum fluctuations. Interesting thoughts for further investigations.

%% file: RP.tex
To conclude, we take a broader perspective and ask how backreaction from quantum processes in dynamical Casimir effect and the like  bear on bigger or deeper issues in problems involving quantum fields interacting with mechanical systems.  We mention two below.

i) The present work is a prelude to a more complex problem using a more thorough approach to {\it quantum optomechanics}. One such model proposed by Galley et al., the so-called mirror-oscillator-field (MOF) model \cite{MOF}, considers the mirror's internal degree of freedom (modeled by a harmonic oscillator) as a dynamical variable interacting with a quantum field. (See also \cite{HarEli,Wang}.)  This provides a microphysics basis for producing different transparencies of the mirror and for a better analysis of mirror-field entanglement \cite{SinLinHu}. The mirror's external or mechanical degree of freedom (dof), i.e., the mirror's position, where the field configuration takes on its value, is also a dynamical variable. With this one can tackle problems involving moving atoms and mirrors and investigate fundamental issues like the Unruh effect and the mirror analog of Hawking effect.   The interplay of these three dynamical variables: the internal dof of the mirror described by an oscillator, the mirror's position, and the quantum field,  takes into account the action and backreaction of one variable on another. Solving these interlinked equations of motions  self-consistently, no doubt a demanding task, gives a more complete quantum description of the optomechanical system. A master equation for the MOF model has been recently derived \cite{ButMOF} which, together with results in a recent paper for two mirrors based on quantum Brownian motion with nonlinear coupling \cite{But2M}, would provide further and better theoretical tools to facilitate more direct applications to problems in quantum optics and nanomechanics. 

ii) In addition to the quantum field  and quantum information theoretical aspects mentioned above, backreaction studies of quantum field processes on mechanical systems also have important quantum thermodynamic implications. Related to the theme of this work one can ask whether a {\it fluctuation-dissipation relation} (FDR) exists in DCE \cite{MazFDR}.  There is hope for this,  because such a relation has been found  in the more complicated problem of the backreaction of cosmological particle creation \cite{CalHu87,HuSin,CamVer}. It conveys the  balance between the total energy of particles created from the vacuum fluctuations of the field and the anisotropy dissipation power integrated over the whole effective period of  particle production.    When one applies an open system viewpoint toward these problems, FDRs are fundamental relations as they serve as a check on the self-consistency of the backreaction problem under study. It is more involved on two counts: a) it is not the  FDR in quantum fields alone, where the causal and Hadamard Green functions are simply related. Rather, what is in analogy with the FDR in cosmological particle creation is a hybrid relation between the parametrically amplified quantum fluctuations in the field and the dissipation in the geometrodynamics (the moving ring or  plates in DCE, the anisotropy or inhomogeneities of the universe). Earlier work on FDR in DCE (e.g. \cite{DNM})  is incomplete in the treatment of noise.  b) one needs to check whether at late times there is a stationary state. For more sophisticated treatments of FDRs in the field (environment) and FDRs in the atom (system) in  nonequilibrium steady state setups, see, e.g., \cite{HH-FDR}. We hope to tackle this problem soon \cite{DCE-FDR}.\\

\noindent {\bf Acknowledgments} 
S.B. acknowledges funding from the Leverhulme Trust Grant No. ECF-2019-461, and from the University of Glasgow via the Lord Kelvin/Adam Smith (LKAS) Leadership Fellowship. B.L.Hu enjoyed the hospitality of the Theory Group at the Institute of Physics, Academia Sinica, Taiwan, ROC, hosted by Prof. Hiang-nan Li, when this work began. They both benefited from participating in the workshop ``Quantum Simulation of Gravitational Problems on Condensed Matter Analog Models'', hosted at the ECT* in Trento, Italy, organized by Prof. Iacopo Carusotto and colleagues.

%% file: appendix1.tex
In this appendix, we present an alternative derivation of the equation of motion for the ring in the 1+1D case, which has the advantage of showing explicitly how the contribution due to the trace anomaly appears. This information is hidden in the effective action we derived in the main text, in Eq.~\eqref{2d action}. To this end, we start from the action for the field-ring system:
\begin{align}
	S = S_a + S_f &= \int \dd t \frac{M}{2}  \dot{a}^2 l^2 + \int \dd t \int_0^l \dd x \;\frac{1}{2}|g|^{1/2}\Big[g^{\mu\nu}\big(\partial_\mu\phi\big)\big(\partial_\nu\phi\big)-m^2\phi^2\Big]\\
 &= \int \dd t \frac{M}{2}  \dot{a}^2 l^2 + \int \dd t \int_0^l \dd x \;\frac{a}{2}\bigg[\big(\partial_t\phi\big)^2-\frac{1}{a^2}\big(\partial_x\phi\big)^2-m^2\phi^2\bigg].
\label{Eq:Action}
\end{align}
In the second line, we used the metric for the time-dependent ring, as introduced in Eq.~\eqref{metric}. The equations of motion for the ring can readily be obtained by minimizing this action with respect to variations in the expansion parameter $a$: $\delta S/\delta a = 0$. By following this procedure, and remembering the formal definition of the energy-momentum tensor of the field
\begin{equation}
	T_{\mu\nu} \equiv \big(\partial_\mu\phi\big)\big(\partial_\nu\phi\big) - \frac{1}{2}g_{\mu\nu} \big(\partial^\rho\phi\big)\big(\partial_\rho\phi\big)+\frac{1}{2}g_{\mu\nu}m^2\phi^2,\\
\label{Eq:EMT}
\end{equation}
we obtain the Euler-Lagrange equation for the expansion parameter in the form:
\begin{equation}
	M\ddot{a}d = -\frac{1}{l} \int_0^l \dd x\, p(x) = -\frac{1}{d} \int_0^l \dd x\, T_1^{\ 1}(x) = \frac{1}{d} \int_0^l \dd x\,\big[T_{00} - T_\rho^{\ \rho}\big].
\label{Eq:ScaleFact2}
\end{equation}
Notice that the momentum density of the field $p(x)\equiv T_1^{\ 1}(x)$ plays the role of a force acting on the ring, and can be written in terms of the energy density $T_{00}$ and the trace $T_\rho^{\ \rho}$. These quantities are the expectation values calculated respect to the quantum state of the field. At this stage, in Eq.~\eqref{Eq:ScaleFact2},   both bare energy density and classical trace appear. They need to be regularized, only their finite  values are physical. We performed this task for the energy density in the main text, and obtained the regularized expression $\rho_{\rm reg} \equiv\expval{T_{00}}_{\rm phy} = -{\pi}/{(6 a^2 l^2)} - {\dot{a}^2}/{(24\pi a^2)}$ (see Eq.~\eqref{Eq:T00_ren}). We now need to regularize the trace $T_\rho^{\ \rho}$. Using the WKB mode function in Eq. \eqref{Eq:WKB}, we can get the expectation value of the trace at the second-adiabatic order. By subtracting this from the bare (divergent) value of the trace, we obtain:
\begin{equation}
\begin{split}
	\expval{T_\rho^{\ \rho}}_{\rm phy} &=  \frac{m^2}{4\pi a}\int_{-\infty}^{+\infty} \dd k\,\left[\frac{1}{W_k}-\left(\frac{1}{W_k}\right)^{(0)}-\left(\frac{1}{W_k}\right)^{(2)}\right]\\
	&= m^2 \big<\phi^2\big>_{\rm reg} + \frac{m^2}{4\pi a}\int_{-\infty}^{+\infty} \dd k\,\frac{W_k^{(2)}}{\omega_k^4}\\
	&= m^2 \big<\phi^2\big>_{\rm reg} + \frac{m^2}{16\pi a}\int_{-\infty}^{+\infty} \dd k\,\frac{1}{\omega_k^3}\left(\frac{3}{2}\frac{\dot{\omega}_k^2}{\omega_k^2}-\frac{\ddot{\omega}_k}{\omega_k}-\frac{1}{2} \left(\frac{\ddot{a}}{a} - \frac{\dot{a}^2}{2a^2}\right)\right). \label{Eq:TAn}
\end{split}
\end{equation}
Taking the massless  limit $m\to 0$ of this expression for a massive field, and by noting that the integral in the second term of \eqref{Eq:TAn} does not depend on the value of $m$ [this can be seen by changing the variable of integral to $k/(ma)$], the trace anomaly takes the simple form:
\begin{equation}
    \expval{T_\rho^{\ \rho}}_{\rm phy} = -\frac{1}{12\pi}\frac{\ddot{a}}{a}.
\end{equation}
This result confirms that the physical value of the trace is equal to $\expval{T_\rho^{\ \rho}}_{\rm phy}=R/(24\pi)$, with $R$ the scalar curvature  of the ring, a geometric invariant, as expected. Substituting the expressions for the regularized energy density and the trace obtained here into  Eq.~\eqref{Eq:ScaleFact2}, we get the equation of motion of the ring:
\begin{equation}
	\left(M - \frac{1}{12\pi L} \right) \ddot{L} = -\frac{\pi}{6 L^2} - \frac{1}{24\pi} \frac{\dot{L}^2}{L^2},\label{EqMot_NoBkr_app}
\end{equation}
which is the same as the one we obtained in the main text, Eq.~\eqref{2d action}, by using the  effective action method.